# SideEye: A Side-Looking Catheter for Fenestrated Endovascular Aneurysm Repair Procedures

**Yara Alawneh[1,4], Alykhan Sewani[1,4],** *Member, IEEE,* **James J. Zhou, Andrew Dueck,[2,3,4] and M. Ali Tavallaei[1,3,4],** *Member, IEEE*

[1]Department of Biomedical Engineering, Faculty of Engineering and Architectural Science, Toronto Metropolitan University
[2]Department of Vascular Surgery, Sunnybrook Health Sciences Centre, Toronto, Ontario, Canada
[3]University of Toronto, Toronto, Ontario, Canada
[4]Sunnybrook Health Sciences Centre, Toronto, Ontario, Canada

CORRESPONDING AUTHOR: M. Ali Tavallaei (e-mail: ali.tavallaei@torontomu.ca)

The authors would like to acknowledge the funding support from the Canada Research Chairs Program (CRC-2019-00012), the Natural Sciences and Engineering Research Council of Canada (NSERC), the Canadian Foundation for Innovation (CFI), the Ontario government, as well as the host academic institutions including Toronto Metropolitan University, and Sunnybrook Health Sciences Centre.

**ABSTRACT** Fenestrated endovascular aneurysm repair remains a technically challenging procedure in the presence of complex anatomy, as it increases the difficulty of target vessel cannulation and prolongs procedure time and fluoroscopy radiation exposure. This paper aims to design, develop, and assess a novel steerable catheter, the SideEye, and compare its performance with conventional catheters in a thoracoabdominal aortic aneurysm phantom model. Users were asked to perform target vessel cannulation under fluoroscopic guidance using the SideEye and conventional non-steerable and steerable catheters. The experiment was divided into two cases based on the stent graft orientation (aligned and misaligned). Total procedure times, individual target vessel cannulation times, and exposure times were analyzed and compared in each case. In the misaligned case, the average cannulation times of all target vessels were 703 ± 274 s using the non-steerable catheter, 517 ± 309 s using the steerable catheter, and 199 ± 91.0 s using the SideEye. The average exposure times were 12 ± 4.6 min using the non-steerable catheter, 8.6 ± 4.1 min using the steerable catheter, and 3.0 ± 1.1 min using the SideEye. Target vessel cannulation using the SideEye significantly reduced procedure time and overall exposure time, compared to conventional devices.

**IMPACT STATEMENT** A side-looking expandable cable driven parallel mechanism is presented which promises to significantly improve the challenging gate cannulation tasks in repair of fenestrated abdominal aortic aneurysms.

## I. INTRODUCTION

A continuous dilation in the thoracic and abdominal segments of the aorta is referred to as a Thoracoabdominal Aortic Aneurysm (TAAA). In this life-threatening condition, the aorta has a high risk of rupture (46% to 74%), if left untreated[5]. Due to the location of the aneurysmal dilation and the involvement of the visceral arteries, TAAA is considered the most complex and challenging aortic aneurysm to treat[5]. Recent advances in stent-graft technology contributed to the treatment of TAAAs, juxtarenal abdominal aortic aneurysms (JAAAs), and suprarenal AAAs by using a minimally invasive endovascular technique known as a Fenestrated Endovascular Aneurysm Repair (FEVAR) procedure[12,21]. This procedure involves the deployment of a four-vessel fenestrated stent graft that extends the seal zone to a healthy segment of the aorta which incorporates the renal or visceral arteries[16].

Although high technical success (>95%) and low mortality rates were reported from large aortic centers worldwide, studies have shown that FEVAR procedural complications were linked with the inability to cannulate the visceral vessels[8,9]. Unsuccessful target vessel cannulation can lead to increased procedure duration, fluoroscopic exposure, contrast volume, and risk of renal failure[9]. Safe cannulation and target vessel preservation is a crucial, challenging, and time-consuming step in a FEVAR procedure[13]. Approximately 40% of patients are denied FEVAR due to unfavorable anatomic characteristics and instead are exposed to the high morbidity and mortality rates associated with an open repair[20]. Additionally, anatomic parameters such as iliac vessel tortuosity, aortic angulation, visceral vessel diameter, length, and angle of incidence to the aorta, and the presence of a stenosis may hinder successful catheterization and increase the risk of end-organ embolization[20].

Furthermore, misalignment of the fenestrated graft with vessel ostia during implantation will result in target vessel cannulation failure[9]. Multiple catheter-guidewire exchanges and cannulation attempts increase the risk of perforation, the strain on blood vessels, and radiation exposure and drastically prolong procedure time[13,22]. Torquability, stability, and flexibility of endovascular devices are essential to successfully cannulate the visceral vessels[11].

Studies have shown that conventional catheters lack the required maneuverability and stability in the presence of access vessel tortuosity and severe angulation of target vessels[7,19]. Alternatively, snaring is a novel endovascular bailout technique that is utilized in challenging situations, where conventional methods fail to cannulate target arteries that are small, tortuous, and angulated downwards[3]. The endovascular snare system is used to apply additional downward force on the delivery sheath to provide stability when cannulating the visceral vessels via transfemoral access. Brachial access can also be used in the snaring technique as it eases cannulation of down-going target vessels, in comparison to femoral access. However, it is associated with an increased risk of stroke, access related complications, and other embolic events[4,6,18].

The "ceiling technique" is another maneuver used when complex target vessel anatomy hinders the cannulation process[17]. It consists of a balloon being inflated just above the target to prevent the guidewire from slipping out of the vessel, due to the tangential component of the pushing force[17]. However, this technique must be implemented with caution as inflating the balloon might crush previously deployed bridging components of other stented target vessels[17].

On the other hand, endovascular electromechanical steerable robotic systems have been shown to be effective in reducing target vessel cannulation times when compared to conventional catheters by providing stability and fine control of catheter/wire tip movements. However, such systems are very costly to implement, generally lack mechanical and haptic feedback—which can have major safety implications—and do not allow the interventionalists to fully apply their dexterous skills[2,14]. Electromagnetic robotic navigation systems also improve stability and maneuverability; however, such systems are costly and require dedicated rooms, customized devices, and the burdensome requirement of continuous checking for magnetic objects in the vicinity of the patient to avoid detrimental effects on accuracy of device navigation as well as for safety reasons[15].

To address these limitations, in this study, we present the design and performance characterization of a novel steerable catheter which we call the SideEye. We also compare the performance of the SideEye to conventional catheters for target vessel cannulation of fenestrated EVAR procedures ex vivo within a TAAA phantom model.

## II. METHODS

The development of the SideEye, as shown in Figure 1, builds on previous work that applied the concept of an e**x**pandable cable-driven parallel mechanism (X-CADPAM) to design a catheter that can be steered with a high level of precision[23]. The steering system is comprised of a user input handle unit, a delivery sheath, an internal catheter, four steering cables, and an expandable frame. The tip of the internal catheter can be steered in a 2D plane by using the handle end to actuate the cables anchored on the distal ends of the expandable frame. For our application in fenestrated EVAR procedures, we present the SideEye, which utilizes a side-looking expandable frame that is suited for deployment inside the stent graft to face the fenestration and an angled internal catheter that can be manipulated to locate the visceral arteries' ostia to then allow for the passage of an interventional device, such as a guidewire, through its lumen to facilitate target vessel cannulation. In the following sections, we describe in detail the design requirements and development process of the proposed system's components.

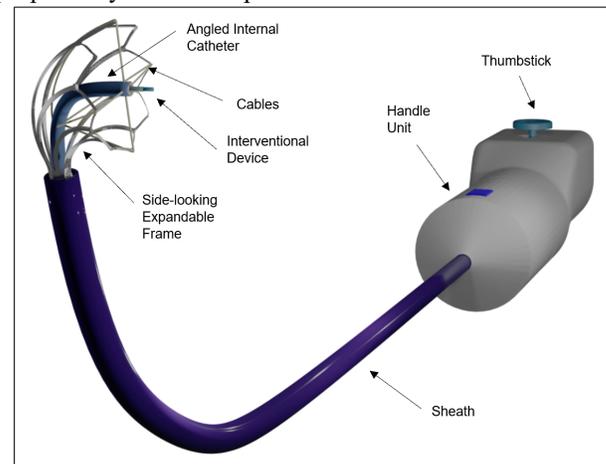

Fig. 1. 3D illustration of the SideEye with its steering system labelled.

**A.** Design Criteria and Design Process
*1) Side-looking Expandable Frame:*
The side-looking expandable frame is a key component in our proposed system. Therefore, to produce a working solution for our specific application, the frame designed had to meet the following requirements:

   I. be able to expand to a diameter of 15 mm;
   I. be side-looking;
  II. be deliverable through a 12Fr sheath;
 III. allow for anchoring of four cables on its distal ends;
 IV. be able to maintain its rigidity when the cables are actuated to steer the internal catheter; and

V. house the angled internal catheter and prevent it from interacting with the walls of the stent graft.

We used a similar design and manufacturing process to that used for a stent, where a specific set of cuts on a cylindrical tube are planned. The tube is then cut with a laser cutting process and formed into shape using various techniques. Using an iterative design approach, we modeled our expandable frame design in SolidWorks (Dassault Systemes, MA, USA, 2021). We assumed a nitinol tube with a wall thickness of 0.2 mm and an ID of 3.0 mm. In our iterative process, various stent cuts were made on the outer surface of the tube to form eight leaflets that are joined together by several connection points. Four of the eight leaflets were chosen as anchor points to the steering cables in the proposed system, and the other four functioned as support structures for mechanical leverage (Figure 1). Further detail on the design iteration process is provided in the Simulations section.

### 2) Angled Internal Catheter:

Device selection is a very crucial step in any interventional procedure as procedural success depends heavily on the ability to torque and manipulate the device[1]. The selection process is based on vessel characteristics, ostial orientation, and degree of support required for device delivery[1]. Therefore, to improve accessibility to visceral vessels ostia and allow for precise placement of interventional devices through its lumen, the catheter designed had to have the following characteristics:
- I) The catheter should have a 90-degree angled tip to easily face the fenestration and locate the target vessels orifices;
- II) The catheter's distal tip should be soft to minimize vessel trauma and dissection;
- III) The catheter's proximal end should be stiff and kink-resistant to allow for pushability and maneuverability of the catheter as it progresses through the body; and
- IV) The catheter's distal segment should be flexible enough to allow for steering but remain firmly engaged while providing mechanical support in the ostium during the introduction of the interventional device through its lumen.

**B.** Simulations

### 1) Frame Design:

To assess the various cut designs on the nitinol tube and to see whether they would provide us with the desired side-looking shape after shape setting, we imported the computer-aided design (CAD) model into ANSYS WB (Ansys, PA, USA, 2021) to simulate the shape-setting process. Using finite element analysis (FEA), we gradually deformed the frame to its desired configuration by using molds of different sizes and shapes (Figure 2A). The following material properties of nitinol[10] were assigned to the stent-cut tube: an elastic modulus of 90 GPa; a yield strength of 1000 MPa; and a Poisson's ratio of 0.3. The molds used to expand and deform the stent-cut tube were assigned the following material properties of structural steel: an elastic modulus of 200 GPa; a yield strength of 250 MPa; and a Poisson's ratio of 0.3. The type of contact interaction between the two bodies was defined as frictionless, in order to simplify the simulation and allow the bodies to slide over each other without resistance. Next, to generate the mesh, adaptive sizing was used to ensure that the appropriate level of resolution was used in each region of the model. This helped reduce the computational load on the simulation, as the mesh size could be adjusted to avoid using unnecessarily fine mesh sizes in regions of the model where they were not needed. Boundary conditions were used to specify the external loads applied to the model. In this case, displacement was defined as the external load to move the cone-shaped fixture 40 mm in the z-axis to expand the stent-cut tube. Constraints, on the other hand, were used to specify how the model is allowed to move and deform under the applied loads and boundary conditions. To prevent the stent-cut tube from moving when the fixture is displaced, a fixed support was applied to its base to prevent any translational or rotational movements in all directions. After setting up the simulation, the solver was run to observe the total deformation of the model in response to the external loads and constraints After several design iterations and simulations, we verified that the side-looking frame design met the specified shape requirements as listed previously above. Once favorable design options for the shape were identified, based on the specified requirements, we continued with further simulations to assess their feasibility as is described in the next section.

### 2) Frame Feasibility:

Further simulations were performed in order to analyze (1) the maximum force required for a 12Fr sheath to collapse the side-looking frame and (2) the force-deflection relationship of the frame when the cables are tensioned with a force of 1.2 N (the maximum force required to steer the internal catheter to the edge of the frame by tensioning the cables)[23]. The side-looking frame was assigned the same nitinol material properties listed previously. The 12Fr delivery sheath was assigned the following material properties of expanded polytetrafluoroethylene[10]: an elastic modulus of 552 MPa; a yield strength of 21,700 MPa; and a Poisson's ratio of 0.3. An explicit dynamics simulation was chosen to simulate the nonlinear dynamics of multibody systems over short time periods. To mesh the geometry, the face sizing feature was used to specify the size and resolution of the finite element mesh in different regions of the model that require a higher level of resolution or where it is expected that there will be a lot of interaction between the two bodies. In terms of boundary conditions, an external load was applied to displace the sheath 30 mm in the z-axis to collapse the side-looking expandable frame. To control the movement and deformation of the model, constraints such as a fixed support were applied to the base of the frame. The solver was then run to observe the model's response to the applied loads and constraints. The Von

Mises stress, under maximum loading conditions, was compared against the yield strength of nitinol to determine the safety factor. In this case, the maximum von Mises stress was 864 MPa, resulting in a safety factor of 1.16 with the yield strength of the nitinol material considered to be 1000 MPa. The simulation results obtained from ANSYS WB demonstrated the feasibility of the design and proved that the expandable side-looking frame, at the size of 15 mm, can smoothly collapse inside a 12Fr sheath with a tolerable force of 24 N (Figure 2B). The setup of the second simulation was very similar to the previous one, but a force was applied to the distal tip instead of displacement. The force applied was 1.2 N and applied to the distal tip of the branch, where the cable is anchored, to determine the corresponding deflection under presumed working conditions. The simulation results estimated that the expandable frame could maintain its rigidity, with negligible deflection (0.26 mm), when the cables are actuated during device operation(Figure 2C).

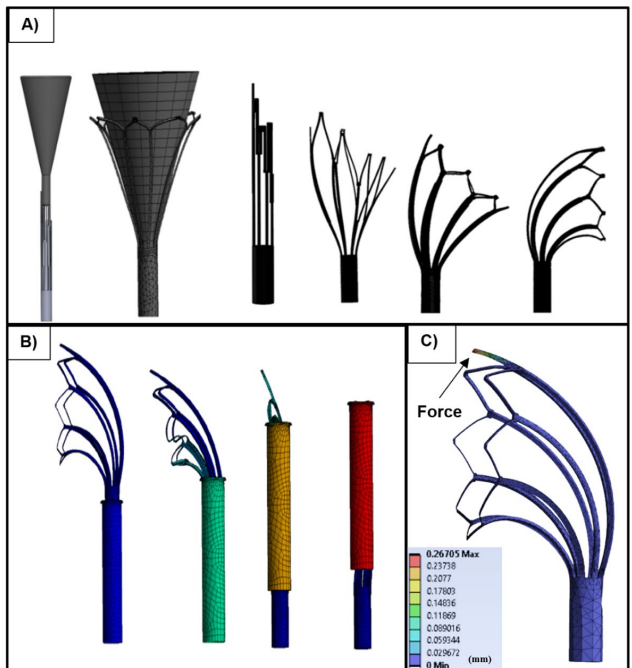

**Fig. 2.** Simulation results A) Incremental expansion and deformation of the nitinol stent-cut tube using molds of different sizes and shapes. B) 12Fr sheath gradually collapsing the side-looking frame by advancing the sheath to its final displacement of 30 mm C) Expandable frame's force-deflection relationship due to cable tensioning.

**C.** Prototype Development
*1) Side-looking Expandable Frame:*
  The side-looking expandable frame was formed from a nitinol tube that was first laser-cut, by means of a fiber laser, to selectively remove material (similar to the manufacturing process of a stent). The laser-cut nitinol tube was then incrementally expanded to its final diameter of 15 mm and deformed to the side by a succession of shape-setting steps that involved heat treatments at 475°F for 20 minutes and the use of custom-designed fixtures to constrain the laser-cut tube in its desired shape (Figure 2A). The frame was then cooled by water quenching and electropolished to round rough edges and smooth out the surface.

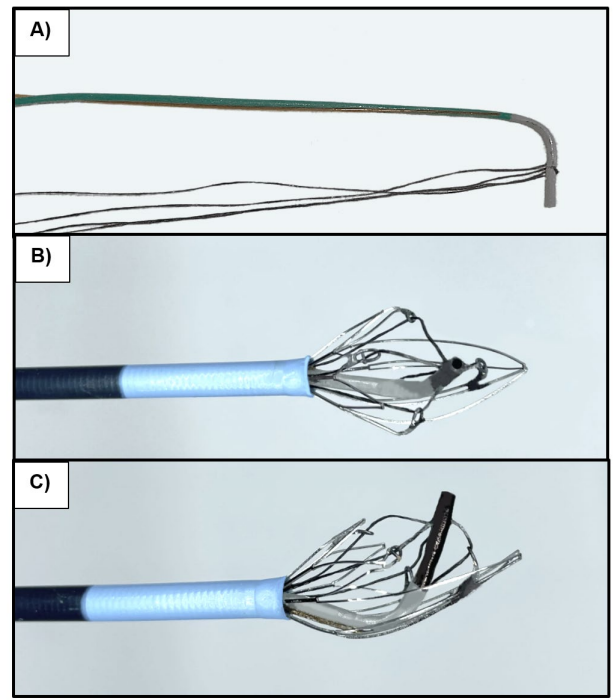

Fig.3. A) An internal catheter with an angled tip. B) A side view of the side-looking expandable frame connected to the angled internal catheter. C) A front view of the assembly.

*2) Angled Internal Catheter:*
  The angled internal catheter, as shown in Figure 3A, was formed from seven main components: a mandrel, base liner, braiding reinforcement, PEBAX sleeves, PTFE cables, polyimide channels, and removable heat shrink. The PTFE liner was placed over a 70 cm mandrel to support the ID of the catheter (1.3 mm) during the construction process and formed a lubricious inner layer with a low coefficient of friction. As part of the X-CADPAM system, four PTFE cables were adhered, 90 degrees apart, to the distal end of the liner to allow for catheter tip manipulation. A heat shrink tube was fitted over a short segment of 40D PEBAX to permit molding it, using a hot air station (Beahm Designs Inc., CA, USA), around the cables to secure them in place. Another segment of 40D PEBAX with fitted heat shrink was melted after the cable connection points to form a soft atraumatic catheter tip. Both distal segments of the mandrel, covered with heat shrink and PEBAX, were deformed while applying heat to form a catheter with a right-angled tip. The mandrel was cooled to preserve the deformed shape, and the heat shrink was then removed. Using a catheter braider (Steeger USA Inc., SC, USA), 304 stainless steel wires were braided over the length of the proximal section of the liner to achieve good torque control and aid in advancing the catheter to the site of interest. The pitch of the braid was modified at each segment to vary the flexibility and stiffness along the catheter shaft. A layer of 72D PEBAX was melted to coat the braided section of the catheter and prevent it from kinking. Four polyimide channels were then glued 90 degrees apart to support the four PTFE cables that run from the catheter end to the handle end.

After completing the construction of each individual component, the expandable frame was attached to the catheter (Figure 3B, 3C). A final layer of 55D PEBAX was melted over the channels and the frame's hypotube to secure the catheter and frame in place. The cables were then routed to anchor on the frame's distal ends and pass through the channels. However, due to the high friction between the braided PTFE cables and the distal tips of the frame, nitinol wires were shape set as eyelets and laser welded on the distal ends to provide a smoother anchoring surface.

**D.** Verification and Validation Experiments

To verify the functionality of the assembled system, the cables exiting the channels were connected to the handle end. By moving the thumbstick on the handle unit (Figure 1) to the four extreme points of the workspace (up/down and left/right), we verified that the angled internal catheter, as designed, can be steered to all four of these points. We also confirmed that the side-looking expandable frame can maintain its rigidity when the cables are actuated. Finally, by advancing a 12Fr sheath over the side-looking expandable frame and retracting it, we verified that the expandable frame is collapsible; we also showed that it maintains its expanding diameter when redeployed (i.e. unconstrained by the sheath).

To validate the SideEye's efficacy in target vessel cannulation and compare its performance with conventional technologies, we designed a TAAA phantom model by obtaining a stereolithography (STL) file generated from a TAAA CT scan that met the criteria for FEVAR (Figure 4A)[24]. The STL file was modified using Autodesk Meshmixer (Autodesk, CA, USA, 2020) to remove artifacts, clean up the mesh, hollow out the model, and create a midline cut at the aneurysm site to allow for manual positioning and rotation of the stent graft (Figure 4B). The modified STL file was then 3D printed using Formlabs clear resin (Formlabs, Somerville, MA, USA) (Figure 4C).

In most fenestrated EVAR procedures, stent grafts with fenestrations are individually customized to each patient based on the location and size of their visceral arteries. However, since custom designed devices were inaccessible for experimenting purposes due to their cost and manufacturing lead time, a Zenith TX2 thoracic stent graft (Cook Medical Inc, Bloomington, IN, USA) was modified to incorporate two small fenestrations (6 mm) and two large fenestrations (10 mm) according to the location of the four visceral arteries on the phantom model. A nitinol ring and radiopaque markers were then sutured around the fenestrations to reinforce the fenestration and enable accurate positioning of the fenestrations relative to their targeted ostia, under fluoroscopy.

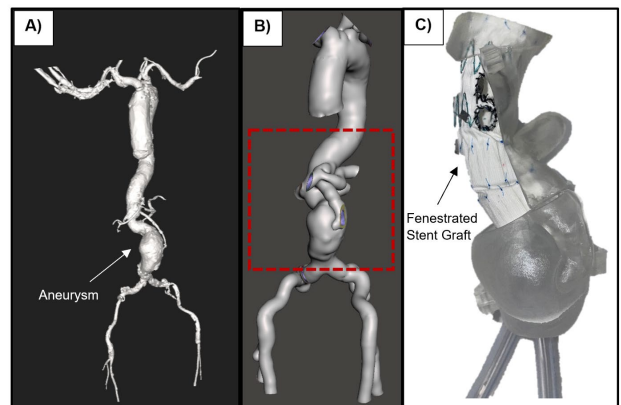

Fig. 4. A) STL file of a thoracoabdominal aortic aneurysm. B) STL file of a TAAA modified using MeshMixer. C) A TAAA phantom model 3D printed using clear resin with the modified stent graft deployed.

**E.** Experimental Protocol

Seven users participated in this study. The users included five novice users (<1 year of experience), one user with moderate experience (>10 years of experience), and one highly experienced vascular interventionalist (>15 years of clinical experience). All were asked to perform target vessel cannulation using the designed phantom to evaluate the performance of the SideEye (Figure 5A) compared to two classes of conventional catheters: non-steerable catheters (GLIDECATH® 4Fr, Terumo Interventional Systems, Tokyo, Japan, Figure 5B), (Cobra 2 IMPRESS® 4Fr, Merit Medical Systems, Utah, USA), (Van Schie 5 Beacon® 5Fr, Cook Medical Inc, Bloomington, IN, USA), (Sos Omni Selective 2 Soft-Vu® 5Fr, AngioDynamics, NY, USA, Figure 5C) and a steerable catheter (AgilisTM NxT 8.5 Fr, Abbott Laboratories, Chicago, IL, USA, Figure 5D). The users were given a selection of non-steerable catheters to use and the freedom to interchange between them as needed. All catheters listed were used to support the steering of a straight 0.035'' or J-shaped guidewire (Glidewire®, Terumo Interventional Systems, Japan). The experiment was divided into two cases: (1) the fenestrations on the stent graft were aligned to directly face the orifices of the target vessels; and (2) the stent graft was rotated (30 degrees) to create a misalignment between the fenestrations and the orifices of the target vessels. For each case, the operators were given a time limit of 15 minutes, starting from when the catheter was located inside the stent graft, to cannulate each target vessel (renal arteries, SMA, and celiac artery), under fluoroscopic guidance, using the SideEye and conventional catheters. The procedure time, radiation dose, and fluoroscopy times of each target vessel cannulation were recorded. Vessel cannulation was considered successful when the guidewire was seen in a stable position and at least 3 cm inside the target vessel. The overall procedural times and fluoroscopy times of all four target vessel cannulations were analyzed for both aligned and misaligned stent graft configurations. Individual target vessel cannulation times and fluoroscopy times were also presented and analyzed for each stent graft configuration (aligned and misaligned).

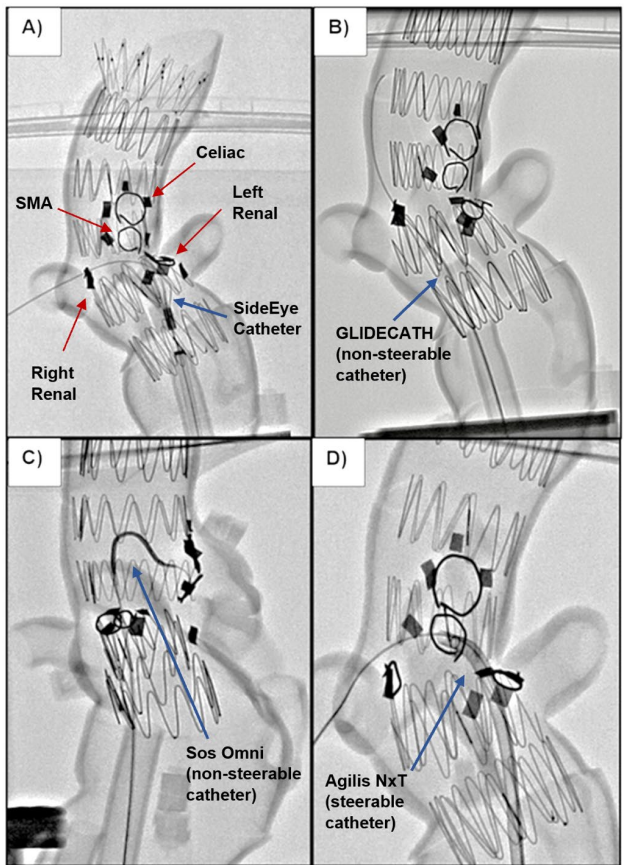

Fig. 5. Fluoroscopic images of target vessel cannulation: A) shows successful cannulation of the right renal artery using the SideEye; B) shows the guidewire deviating upward using a non-steerable catheter (GLIDECATH), in the case of a misaligned stent graft; C) shows buckling of the guidewire using a Sos Omni Selective 2 catheter; and D) shows the steerable catheter cannulating the right renal artery when attempting to cannulate the superior mesenteric artery.

PRISM software (GraphPad) was used for data analysis. Differences between SideEye and conventional vessel cannulation were statistically compared using a two-way analysis of variance (ANOVA) for parametric datasets and using the Kruskal-Wallis test for non-parametric datasets. A $p$ value below 0.05 was considered statistically significant.

### III. RESULTS

#### A. Overall Procedure

*Case A.1) Aligned Stent Graft:*
In an aligned stent graft configuration, all users achieved success with all catheters when cannulating the four target vessels. However, there were statistically significant differences in total cannulation times between the non-steerable catheter, the steerable catheter and the SideEye ($p=0.0149$, Kruskal-Wallis test, Figure 6A). The average times for cannulation of all target vessels were 510 ± 337 s using the non-steerable catheter, 509 ± 397 s, using the steerable catheter, and 177 ± 74.4 s using the SideEye. The cannulation times between novice and expert users did not significantly differ for the three methods (*p=0.121*). We further analyzed the total cannulation times with post-hoc tests and found a significant statistical difference between the SideEye and the non-steerable catheter (*p=0.0331*), but we were unable to find a statistical difference between the SideEye and the steerable catheter (*p=0.0753*) for the aligned stent graft case.

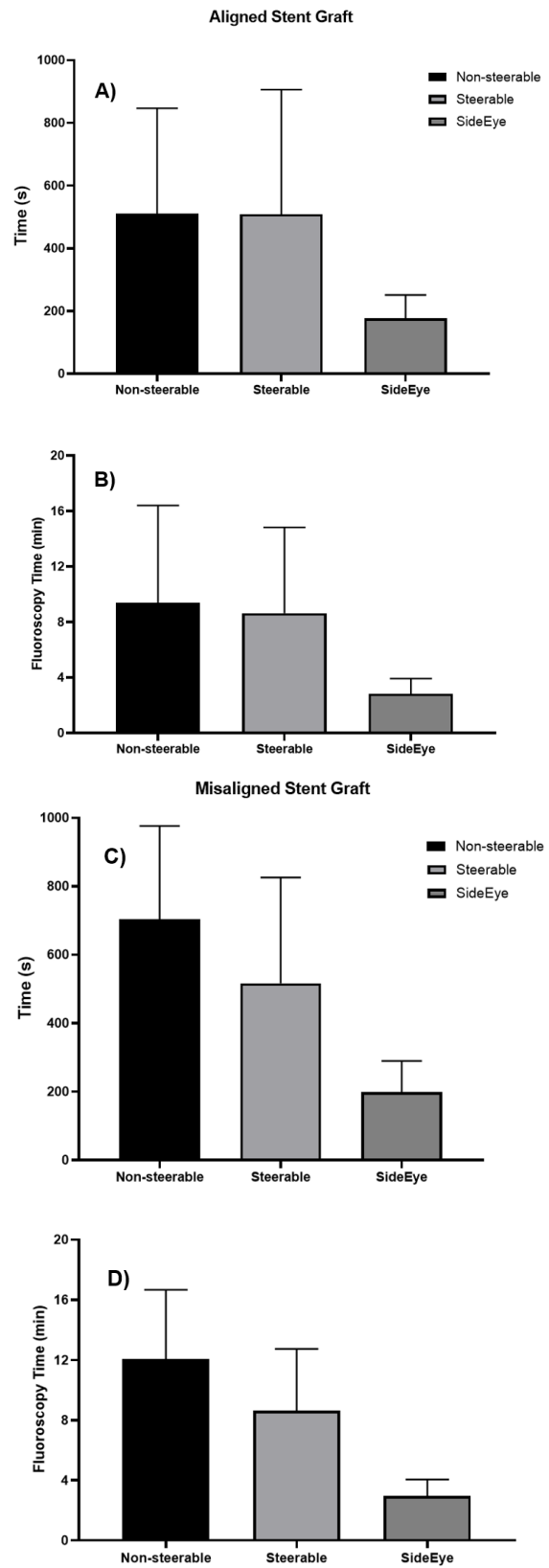

Fig. 6. Cannulation results of all target vessels in the TAAA phantom model using a non-steerable catheter, a steerable catheter, and the SideEye. A) shows the cannulation times in an aligned stent graft configuration; B) shows the fluoroscopy times in an aligned stent graft configuration; C) illustrates the cannulation times in a misaligned configuration; and D) illustrates the fluoroscopy times in a misaligned configuration.

For total fluoroscopy times, there were significant differences between the conventional catheters and the SideEye (*p=0.0046,* Kruskal-Wallis test, Figure 6B). Using the non-steerable catheter, the steerable catheter, and the SideEye, average exposure times were 9.4 ± 7.0 min, 6.2 ± 2.3 min, and 2.8 ± 1.1 min, respectively. After further analysis using post-hoc tests, we found a significant statistical difference in fluoroscopy times between the SideEye versus the non-steerable catheter (*p=* 0.0153) and the SideEye versus the steerable catheter (*p=0.0374*).

*Case A.2) Misaligned Stent Graft:*

In the case of a misaligned stent graft orientation, cannulations of all four target vessels were successful using the conventional catheters and the SideEye. However, there were statistically significant differences in cannulation times between the three different catheters (*p=0.0003,* Kruskal-Wallis test, Figure 6C). The average times for cannulation of all target vessels using the non-steerable catheter, the steerable catheter, and the SideEye were 703 ± 274 s, 517 ± 309 s, and 199 ± 91.0 s, respectively. The cannulation times between novice and expert users did not significantly differ for the three methods (*p=0.0777*). We also further analyzed the total cannulation times with post-hoc tests and found a significant statistical difference between the SideEye and the non-steerable catheter (*p=* 0.0017), but we were unable to observe a statistical difference between the SideEye and the steerable catheter (*p=* 0.0535).

For total fluoroscopy times, there were significant differences between the conventional catheters (grouped) and the SideEye (*p=0.0001,* Kruskal-Wallis test, Figure 6D). The average exposure times recorded from the C-arm were 12 ± 4.6 min using the non-steerable catheter, 8.6 ± 4.1 min using the steerable catheter, and 3.0 ± 1.1 min using the SideEye. After further analysis using post-hoc tests, we found a statistically significant difference between the SideEye versus the non-steerable catheter (*p=0.001*) and the SideEye versus the steerable catheter (*p=0.0274*) in terms of exposure time.

B. *Individual Target Vessel Cannulation*

*Case B. 1) Aligned Stent Graft:*

In an aligned stent graft configuration, no statistically significant difference was found in cannulation times between the four target vessels when the data of three methods were grouped together (*p=0.1284,* two-way ANOVA, Figure 7A). However, cannulation times were significantly different between the three methods (*p=0.0045*). The average times for cannulation of the superior mesenteric artery were 136 ± 92.3 s using the non-steerable catheter, 65.9 ± 41.8 s using the steerable catheter, and 29.0 ± 16.7 s using the SideEye. As seen in Figure 7A, the average times to cannulate the left renal artery using the non-steerable catheter, the steerable catheter, and the SideEye were 153 ± 104 s, 202 ± 183 s, and 61.3 ± 23.3 s, respectively. After further analysis of the cannulation times using post-hoc tests, a significant difference was found for the SideEye versus the non-steerable catheter (*p=-0.0114*), and the SideEye versus the steerable catheter (*p=0.0117*).

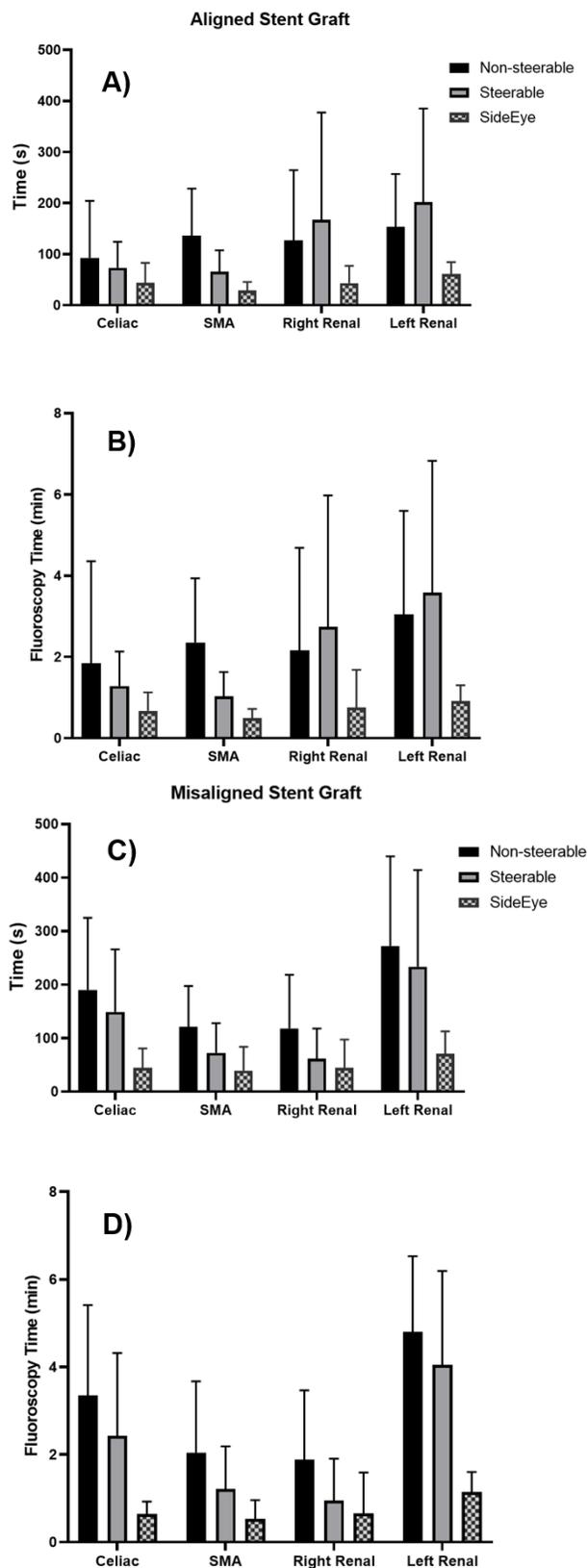

Fig. 7. Cannulation results of each individual target vessel using the non-steerable catheter, the steerable catheter, and the SideEye. A) shows cannulation times of each target vessel, in an aligned stent graft configuration; B) shows fluoroscopy times in an aligned stent graft configuration; C) illustrates cannulation times of each target vessel in a misaligned stent graft configuration; and D) illustrated fluoroscopy times in a misaligned configuration.

For fluoroscopy times, there were no significant differences between the four target vessels (*p=0.1285, two-way ANOVA, Figure 7B*). However, there were significant differences in exposure times between the conventional catheters and the SideEye (*p=0.0038*). When cannulating the superior mesenteric artery, average exposure times for ionizing radiation were 2.4 ± 1.6 min using the non-steerable catheter, 1.0 ± 0.60 min using the steerable catheter, and 0.49 ± 0.23 min using the SideEye. During the cannulation of the left renal artery using the non-steerable catheter, the steerable catheter, and the SideEye average exposure times (Figure 7B) were 3.0 ± 2.6 min, 3.6 ± 3.2 min, and 0.91 ± 0.39 min, respectively. After further analysis of fluoroscopy times using post-hoc tests, we found a significant difference between the SideEye versus the non-steerable catheter (*p=0.0062*) and the SideEye versus the steerable catheter (*p=0.0174*).

*Case B.2) Misaligned Stent Graft:*
In a misaligned stent graft configuration, there were significant differences in cannulation times between the four target vessels (*p=0.0007, two-way ANOVA, Figure 7C*) and the three different methods (*p=0.0001*). The average times for cannulation of the celiac artery were 190 ± 134 s using the non-steerable catheter, 149 ± 117 s using the steerable catheter, and 44.3 ± 36.7 s using the SideEye. As shown in Figure 7C, the average times to cannulate the left renal artery using the non-steerable catheter, the steerable catheter, and the SideEye were 272 ± 167 s, 233 ± 181 s, and 70.9 ± 41.8 s, respectively. After further analysis of cannulation times using post-hoc tests, a significant difference was found between the SideEye versus the non-steerable catheter (*p=0.0001*) , and the SideEye versus the steerable catheter (*p= 0.0113*).

There were also very significant differences in fluoroscopy times between the four target vessels (*p= 0.0001, two-way ANOVA, Figure 7D*), and the three different methods used (*p=0.0001*). To cannulate the celiac artery average exposure times were 3.3 ± 2.1 min using the non-steerable catheter, 2.4 ± 1.9 min using the steerable catheter, and 0.64 ± 0.28 min using the SideEye. During the cannulation of the left renal artery, average exposure times (Figure 7D) were 4.8 ± 1.7 min, 4.0 ± 2.1 min, and 1.1 ± 0.46 min using the non-steerable catheter, the steerable catheter, and the SideEye, respectively. Again, after further analysis of radiation exposure times using post-hoc tests, a significant difference was found between the SideEye versus the non-steerable catheter (*p=0.0001*), and the SideEye versus the steerable catheter (*p=0.001*).

## IV. DISCUSSION

Target vessel cannulation through fenestrations can be made challenging by tortuous anatomy, small arteries, misaligned fenestrations with target vessel orifices, and downward angulation of the arteries. Standard techniques using conventional catheters in unpredictable anatomy, may be limited in terms of catheter pushability, maneuverability, torqueability, and stability. As a result, adequate catheter positioning and guidewire support at the vessel ostia remains a frequent problem in target vessel cannulation.

To address these limitations, we developed a novel steerable catheter, the SideEye, to stabilize target vessel cannulation and allow for precise catheter tip manipulation in challenging anatomy during fenestrated EVAR procedures. In this study, we quantified and compared the performance of the SideEye in target vessel cannulation to conventional non-steerable and steerable catheters using a 3D printed model of a thoracoabdominal aortic aneurysm.

In the case of an aligned stent graft, cannulation of all target vessels using the SideEye was significantly faster than standard cannulation using conventional non-steerable catheters. The average exposure time when cannulating all four vessels was significantly reduced using the SideEye compared to the non-steerable and the steerable catheter. In the case of a misaligned stent graft configuration, cannulations of the celiac artery, and the left renal artery were significantly faster using the SideEye compared to non-steerable and steerable catheters. Moreover, when cannulating the left renal artery specifically, exposure times were significantly lower using the SideEye compared to the steerable and non-steerable catheters. The SideEye was successfully used for cannulating the target vessels of the phantom by utilizing the side-looking expandable frame to face the fenestration and provide a stable platform for the passage of the guidewire through the orifices of the target vessel. The conventional non-steerable catheter, although successful at cannulation after a few attempts, tends to deviate upwards towards the thoracic aorta after the introduction of a guidewire through its lumen, leading to traumatic manipulations. Although operators had the option to change between standard non-steerable catheters and catheters with complex shapes like the Sos Omni Selective 2 catheter to provide precise guidewire manipulation, angulation at the access point has limited stability and resulted in buckling of the guidewire. While no statistical difference was found in total cannulation times between the SideEye and the steerable catheter, attempts to cannulate the superior mesenteric artery using the steerable catheter resulted in cannulation of the right renal artery due to difficulties in precise positioning and steering of the sheath from inside the stent graft.

This study is limited by the use of an ex vivo phantom model for target vessel cannulation as it does not accurately reflect the challenges encountered in a clinical setting. However, as part of future studies, we plan to explore the safety of using the proposed system in vivo and assess its feasibility in fenestrated EVAR procedures. Another limitation is that most users were inexperienced in catheter manipulation. However, the comparison of cannulation performance between novice and expert users revealed that the results by the expert users mirror the trends seen by the group as a whole.

V. CONCLUSION

In conclusion, in this study we have proposed a novel side-looking steerable catheter that utilizes an expandable cable-driven mechanism. Within a 3D printed aneurysm model, we were able to show that the SideEye significantly outperforms conventional devices for visceral artery cannulation in terms of procedure time and overall radiation exposure time.

VI. REFERENCES

1. Ali, R., A. B. Greenbaum, and A. D. Kugelmass. A Review of Available Angioplasty Guiding Catheters, Wires and Balloons – Making the Right Choice. *Interventional Cardiology Review* 7:100, 2012.
2. Dagnino, G., J. Liu, M. E. M. K. Abdelaziz, W. Chi, C. Riga, and G. Z. Yang. Haptic Feedback and Dynamic Active Constraints for Robot-Assisted Endovascular Catheterization. *IEEE International Conference on Intelligent Robots and Systems* 1770–1775, 2018.doi:10.1109/IROS.2018.8593628
3. Ferreira, M., A. Katsargyris, E. Rodrigues, D. Ferreira, R. Cunha, G. Bicalho, G. Oderich, and E. L. G. Verhoeven. "snare-Ride": A Bailout Technique to Catheterize Target Vessels with Unfriendly Anatomy in Branched Endovascular Aortic Repair. *Journal of Endovascular Therapy* 24:556–558, 2017.
4. Kret, M. R., R. L. Dalman, J. Kalish, and M. Mell. Arterial cutdown reduces complications after brachial access for peripheral vascular intervention. *J Vasc Surg* 64:149–154, 2016.
5. Lobato, A. C., and L. Camacho-Lobato. A New Technique to Enhance Endovascular Thoracoabdominal Aortic Aneurysm Therapy-The Sandwich Procedure. , 2012.doi:10.1053/j.semvascsurg.2012.07.005
6. Madden, N. J., K. D. Calligaro, H. Zheng, D. A. Troutman, and M. J. Dougherty. Outcomes of Brachial Artery Access for Endovascular Interventions. *Ann Vasc Surg* 56:81–86, 2019.
7. Mazzaccaro, D., E. L. Castronovo, P. Righini, and G. Nano. Use of steerable catheters for endovascular procedures: Report of a CASE and literature review. *Catheterization and Cardiovascular Interventions* 95:971–977, 2020.
8. Oderich, G. S., M. Ribeiro, J. Hofer, J. Wigham, S. Cha, J. Chini, T. A. Macedo, and P. Gloviczki. Prospective, nonrandomized study to evaluate endovascular repair of pararenal and thoracoabdominal aortic aneurysms using fenestrated-branched endografts based on supraceliac sealing zones. *J Vasc Surg* 65:1249-1259.e10, 2017.
9. Pena, C. S., B. J. Schiro, and J. F. Benenati. Fenestrated Endovascular Abdominal Aortic Aneurysm Repair. *Tech Vasc Interv Radiol* 21:156–164, 2018.
10. Randhawa, H. S., G. Pearce, R. Hepton, J. Wong, I. F. Zidane, and X. Ma. An investigation into the design of a device to treat haemorrhagic stroke. *Proc Inst Mech Eng H* 234:323–336, 2020.
11. Riga, C. v., C. D. Bicknell, M. Hamady, and N. Cheshire. Tortuous Iliac Systems—A Significant Burden to Conventional Cannulation in the Visceral Segment: Is There a Role for Robotic Catheter Technology? *Journal of Vascular and Interventional Radiology* 23:1369–1375, 2012.
12. Riga, C. v, C. D. Bicknell, M. Hamady, and N. Cheshire. Tortuous Iliac Systems-A Significant Burden to Conventional Cannulation in the Visceral Segment: Is There a Role for Robotic Catheter Technology?doi:10.1016/j.jvir.2012.07.006
13. Riga, C. v., C. D. Bicknell, A. Rolls, N. J. Cheshire, and M. S. Hamady. Robot-assisted Fenestrated Endovascular Aneurysm Repair (FEVAR) Using the Magellan System. *Journal of Vascular and Interventional Radiology* 24:191–196, 2013.
14. Riga, C. v., N. J. W. Cheshire, M. S. Hamady, and C. D. Bicknell. The role of robotic endovascular catheters in fenestrated stent grafting. *J Vasc Surg* 51:810–820, 2010.
15. de Ruiter, Q. M. B., F. L. Moll, and J. A. van Herwaarden. Current state in tracking and robotic navigation systems for application in endovascular aortic aneurysm repair. *J Vasc Surg* 61:256–264, 2015.
16. Schwein, A., P. Chinnadurai, G. Behler, A. B. Lumsden, J. Bismuth, and C. F. Bechara. Computed tomography angiography-fluoroscopy image fusion allows visceral vessel cannulation without angiography during fenestrated endovascular aneurysm repair. *J Vasc Surg* 68:2–11, 2018.
17. Simonte, G., G. Isernia, G. Fino, L. Baccani, A. Fino, F. Casali, and M. Lenti. Ceiling Technique to Facilitate Target Vessel Catheterization During Complex Aortic Repair. *Ann Vasc Surg* 71:528–532, 2021.
18. Swerdlow, N. J., P. Liang, C. Li, K. Dansey, T. F. X. O'Donnell, L. E. V. M. de Guerre, R. R. B. Varkevisser, V. I. Patel, G. J. Wang, and M. L. Schermerhorn. Stroke rate after endovascular aortic interventions in the Society for Vascular Surgery Vascular Quality Initiative. *J Vasc Surg* 72:1593–1601, 2020.
19. Tangen, G. A., F. Manstad-Hulaas, E. Nypan, and R. Brekken. Manually Steerable Catheter With Improved Agility. *Clin Med Insights Cardiol* 12:, 2018.
20. Timaran, C. H., G. A. Stanley, M. S. Baig, D. E. Timaran, J. G. Modrall, and M. Knowles. The sequential catheterization amid progressive endograft deployment technique for fenestrated endovascular aortic aneurysm repair. *J Vasc Surg* 66:311–315, 2017.
21. Verhoeven, E. L. G., G. Vourliotakis, W. T. G. J. Bos, I. F. J. Tielliu, C. J. Zeebregts, T. R. Prins, U. M. Bracale, and J. J. A. M. van den Dungen. Fenestrated Stent

Grafting for Short-necked and Juxtarenal Abdominal Aortic Aneurysm: An 8-Year Single-centre Experience. *European Journal of Vascular and Endovascular Surgery* 39:529–536, 2010.
22. Walker, T. G., S. P. Kalva, K. Yeddula, S. Wicky, S. Kundu, P. Drescher, B. J. D'Othee, S. C. Rose, and J. F. Cardella. Clinical Practice Guidelines for Endovascular Abdominal Aortic Aneurysm Repair: Written by the Standards of Practice Committee for the Society of Interventional Radiology and Endorsed by the Cardiovascular and Interventional Radiological Society of Europe and the Canadian Interventional Radiology Association. *Journal of Vascular and Interventional Radiology* 21:1632–1655, 2010.
23. Zhou, J. J., A. Quadri, A. Sewani, Y. Alawneh, R. Gilliland-Rocque, C. Magnin, A. Dueck, G. A. Wright, and M. A. Tavallaei. The CathPilot: A Novel Approach for Accurate Interventional Device Steering and Tracking. *IEEE/ASME Transactions on Mechatronics* 1–12, 2022.doi:10.1109/TMECH.2022.3188955
24. File: Aortic Aneurism 76F 3D SR Nevit Dilmen.stl - Wikimedia Commonsat <https://commons.wikimedia.org/wiki/File:Aortic_Aneurism_76F_3D_SR_Nevit_Dilmen.stl#/media/File:Aortic_Aneurism_76F_3D_SR_Nevit_Dilmen.stl>